\documentclass[letterpaper]{article} 
\usepackage[preprint]{aaai2027}  
\usepackage[hyphens]{url}  
\usepackage{graphicx} 
\usepackage{natbib}  
\usepackage{caption} 
\usepackage{algorithm}
\usepackage{algorithmic}

\usepackage{newfloat}
\usepackage{listings}
\DeclareCaptionStyle{ruled}{labelfont=normalfont,labelsep=colon,strut=off} 
\floatstyle{ruled}
\newfloat{listing}{tb}{lst}{}
\floatname{listing}{Listing}

\usepackage{booktabs}

\usepackage{amssymb}
\usepackage{amsmath}
\usepackage{comment}

\title{MorphAtt: A Neuromorphic Accelerator for Efficient Multi-Head Attention Processing in Spiking Vision Transformers}
\author{
    Rachmad Vidya Wicaksana Putra$^*$,
    Amirhesam Jafari Rad$^{\dagger}$,
    Muhammad Shafique$^*$
}
\affiliations{
    $^*$eBRAIN Lab, New York University (NYU) Abu Dhabi, Abu Dhabi, United Arab Emirates \\
    $^{\dagger}$University of Tehran, Tehran, Iran \\
    \{rachmad.putra, muhammad.shafique\}@nyu.edu, a.h.jafarirad@gmail.com 
}

\begin{document}

\maketitle

\begin{abstract}
Spiking Vision Transformers (SViTs) are developed as an energy-efficient alternative to conventional ViTs for computer vision tasks at the edge.
However, huge parameter counts and complex multi-head self-attention (MHSA) operations make it challenging to achieve high energy efficiency in SViT inference, especially in tightly constrained applications.
To maximize efficiency gains of SViT processing, we propose \textit{\textbf{MorphAtt}}, a novel digital accelerator that expedites SViT inference through streamlined processing.
Specifically, it processes MHSA operations using cascaded hardware modules: a Spiking Query-Key-Value generator (SpikeQKV), a low-complexity Spiking Multi-Head Self-Attention engine (SpikeAtten), and Reparameterization Convolution  (RepConv) modules.
To mitigate traffic congestion in on-chip memory accesses and data reuse, specialized inter-module buffers are integrated within the dataflow.  
Under synthesis using 32nm CMOS technology, MorphAtt achieves 792-1605 GOPS of throughput, while incurring $\sim$39-55 mW of power consumption and 1.5 mm$^2$ of area, which lead to 20.3-29.1 TOPS/W of energy efficiency. 
These results also demonstrate that our MorphAtt offers better performance and efficiency trade-offs than state-of-the-art, thereby enabling highly energy-efficient vision-based AI systems at the edge.
\end{abstract}


\section{Introduction}
\label{Sec_Intro}

Vision Transformers (ViTs) have outperformed conventional artificial neural networks (ANNs) in solving various computer vision-based tasks, hence becoming the state-of-the-art algorithms~\cite{Ref_Dosovitskiy_Transformers_ICLR21, Ref_Touvron_TrainingDeIT_ICML21, Ref_Khan_SurveyViT_CSUR22, Ref_Han_SurveyViT_TPAMI22}.
However, standard ViTs rely on a huge number of parameters and complex multi-head self-attention (MHSA) operations to achieve state-of-the-art accuracy, thereby hindering their energy-efficient deployments for tightly constrained applications. 
To improve the efficiency gains of ViT inference, alternative ViT models based on event-driven operations from Spiking Neural Networks (SNNs) have been developed~\cite{Ref_Zhou_Spikformer_ICLR23, Ref_Yao_SpikeDrivenTransformer_NeurIPS23, Ref_Yao_SpikeDrivenTransformer2_ICLR24}, so-called \textit{Spiking Vision Transformers (SViTs)}.
SViTs replace energy-hungry floating-point matrix multiplications with energy-efficient spike-based operations.
Despite offering lower energy consumption than conventional ViTs, the huge number of parameters in SViTs still poses performance and efficiency challenges due to high latency and energy consumption of memory access and computation in SViT processing.  
\textit{Therefore, the \textbf{targeted research problem} in this paper is: how can we effectively achieve high performance and efficiency gains for SViT inference?}

\begin{figure}[t]
    \centering
    \includegraphics[width=\linewidth]{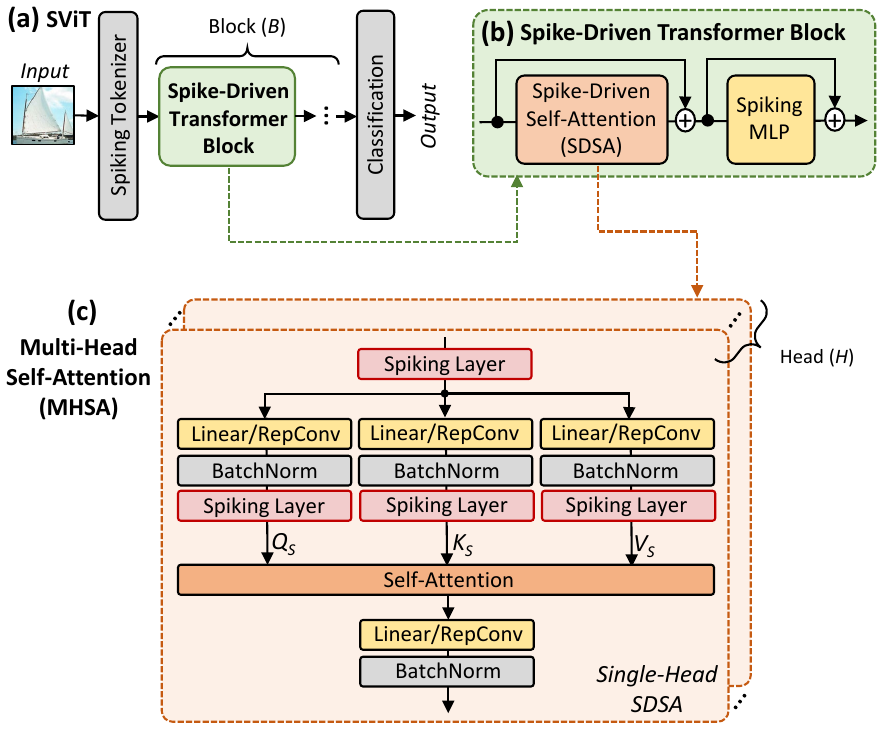}
    \caption{A typical SViT architecture: \textbf{(a)} a network overview, \textbf{(b)} spike-driven transformer block, and \textbf{(c)} multi-head self-attention (MHSA).}
    \label{Fig_MHSA}
    \vspace{-0.2cm}
\end{figure}

To address this problem, specialized digital hardware accelerators have been proposed in the literature~\cite{Ref_Chen_VESTA_APCCAS24, Ref_Xu_SThw3D_ICCAD24, Ref_Gao_SpikeTA_TCAD25, Ref_Li_FireFlyT_arXiv25, Ref_Xu_Bishop_ISCA25, Ref_Fang_EMAfree_TCASAI25}. 
While these state-of-the-art designs demonstrate improvements in execution speed, they still have several limitations.
First, existing FPGA-based designs often incur relatively high power consumption during operation (e.g., more than 4W), while dedicated ASIC designs still consume over 200mW. 
Such high power consumption makes them unsuitable for applications with stringent power requirements and long battery life expectations (e.g., mobile autonomous agents/robots).
Second, they often rely on global on-chip memory to store data and parameters during the processing, which leads to high traffic congestion in on-chip memory access, and hence power/energy consumption.
Third, they have not considered state-of-the-art SViT models, such as Spike-Driven Transformer v2 (SDTv2)~\cite{Ref_Yao_SpikeDrivenTransformer2_ICLR24}, which employ reparameterization modules to improve their inference performance, as shown in Figure~\ref{Fig_MHSA}.
\textit{These limitations highlight the need for an alternative ultra-low power/energy SViT accelerator.}
However, designing such an accelerator is non-trivial due to the complex nature of MHSA operations.

To address this limitation, we propose \textit{\textbf{MorphAtt}}, a novel digital ASIC accelerator targeting low-power SViT inference. 
MorphAtt leverages a cascaded design with buffers between modules to optimize the MHSA and reparameterization modules on-chip; see an overview in Figure~\ref{Fig_MorphAtt}.
Through our MorphAtt design, we provide the following \textbf{novel contributions}. 
\begin{enumerate}
    \item We design \textbf{a streamlined hardware architecture} that executes MHSA operations through cascaded hardware modules including a Spiking Query-Key-Value generator (SpikeQKV), a low-complexity Spiking Multi-Head Self-Attention engine (SpikeAtten), and Reparameterization Convolution (RepConv) modules with inter-module buffers to prevent a memory access bottleneck.
    \item We develop \textbf{an efficient Spiking Multi-Head Self-Attention engine (SpikeAtten)} by employing an optimized processing dataflow that reorders Query-Key-Value operations, so that the computation complexity is reduced from $O(N^2D)$ to $O(ND^2)$; with $N$ is the number of input tokens and $D$ is the embedding dimension. 
    \item We perform \textbf{a comprehensive design synthesis} considering the 32nm CMOS technology, and our MorphAtt achieves 792-1605 GOPS throughput, while consuming $\sim$39-55 mW power and 1.5 mm$^2$ area, leading to 20.3-29.1 TOPS/W and offering competitive performance and efficiency trade-offs compared to the state-of-the-art, making it suitable for edge-AI applications. 
\end{enumerate}

\section{Related Work}
\label{Sec_Related}

\subsection{Spiking Neural Networks (SNNs)}
\label{Sec_Related_SNNs}

SNNs are widely accepted as the third generation of NN computation model due to their bio-plausible design approach~\cite{Ref_Maass_SNN_NeuNet97}.
An SNN model is formed based on the combination of a spiking neuron model, a network topology/architecture, a learning rule, and a neural/spike coding scheme~\cite{Ref_Putra_FSpiNN_TCAD20}.
SNNs typically employ a leaky integrate-and-fire (LIF) neuron model due to their event-based data representation and computation~\cite{Ref_Tavanaei_DLSNN_Neunet18, Ref_Pfeiffer_DLSNN_FNINS18, Ref_Putra_ReSpawn_ICCAD21, Ref_Putra_RescueSNN_FNINS23, Ref_Putra_SoftSNN_DAC22}.

\subsection{Spiking Vision Transformers (SViTs)}
\label{Sec_Related_SViTs}
SNNs have shown their potential as an energy-efficient alternative to neural network algorithms~\cite{Ref_Akopyan_TrueNorth_TCAD15, Ref_Davies_Loihi_MM18, roy2019towards, Ref_Rathi_SNNsurvey_CSUR23, Ref_Putra_SNNonCNP_IJCNN25, Ref_Putra_SpikeNAS_TAI25}. 
Therefore, researchers have been actively leveraging SNN-based components and operations for vision transformers, so-called \textit{Spiking Vision Transformers (SViTs)}.
Several state-of-the-art SViT models have been proposed in the literature, including Spikformer \cite{Ref_Zhou_Spikformer_ICLR23}, 
Spike-Driven Transformer (SDT)~\cite{Ref_Yao_SpikeDrivenTransformer_NeurIPS23}, 
SpikingResformer~\cite{Ref_Shi_SpikingResformer_CVPR24}, 
and Spike-Driven Transformer v2 (SDTv2)~\cite{Ref_Yao_SpikeDrivenTransformer2_ICLR24}. 
Despite different overall network architectures, these models essentially share similar MHSA structures, including spiking convolutional (CONV) layers, spiking linear layers, and spiking attention layers; as shown in Figure~\ref{Fig_MHSA}(c). 

\subsection{SViT Accelerators}
\label{Sec_Related_SViTacc}

Recently, several digital hardware accelerators for SViT inference have been proposed in the literature~\cite{Ref_Chen_VESTA_APCCAS24, Ref_Xu_SThw3D_ICCAD24, Ref_Gao_SpikeTA_TCAD25, Ref_Li_FireFlyT_arXiv25, Ref_Xu_Bishop_ISCA25, Ref_Fang_EMAfree_TCASAI25}. 
They can be loosely classified into several categories based on their implementation paradigm, such as FPGA-based accelerators~\cite{Ref_Li_SViTacc_APACE24, Ref_Gao_SpikeTA_TCAD25, Ref_Li_FireFlyT_arXiv25}, 2D ASIC-based accelerators~\cite{Ref_Chen_VESTA_APCCAS24, Ref_Xu_Bishop_ISCA25, Ref_Fang_EMAfree_TCASAI25}, and 3D ASIC-based accelerators~\cite{Ref_Xu_SThw3D_ICCAD24}. 
In general, FPGA-based designs incur relatively high power consumption (i.e., more than 4W) during operation due to the nature of gate-array programming. 
Meanwhile, dedicated ASIC-based accelerators typically consume over 200mW of power. 
These power profiles make existing solutions unsuitable for edge-AI applications with stringent power budgets and long battery life expectations (e.g., lightweight mobile autonomous agents/robots).

\section{Our \textit{MorphAtt} Accelerator}
\label{Sec_Accel}

\begin{figure*}[t]
    \centering
    \includegraphics[width=\linewidth]{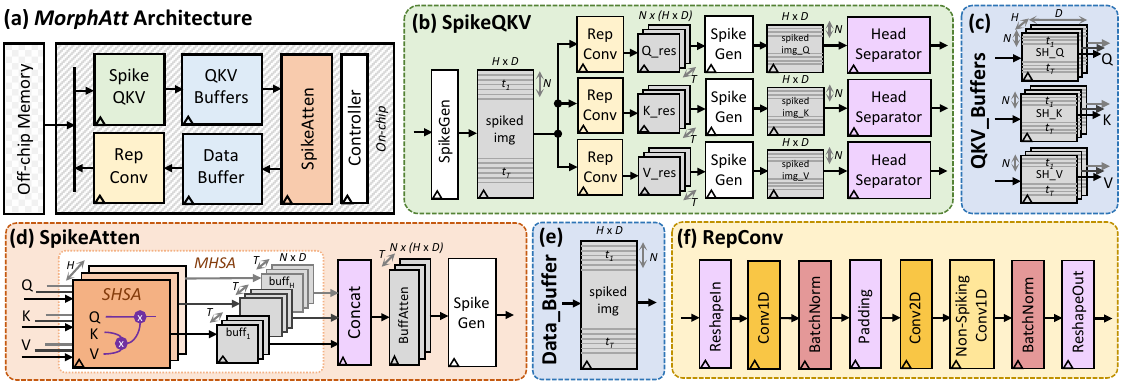}
    \caption{Our MorphAtt accelerator: \textbf{(a)} the overall architecture, and the detailed architectures for key modules: \textbf{(b)} Spiking Query-Key-Value generator (SpikeQKV), \textbf{(c)} QKV buffers (QKV\_Buffers), \textbf{(d)} Spiking Multi-Head Self-Attention engine (SpikeAtten), \textbf{(e)} data buffer (Data\_Buffer), and \textbf{(f)} Reparameterization Convolution module (RepConv).}
    \label{Fig_MorphAtt}
\end{figure*}

To address the targeted research problem and the limitations of state-of-the-art SViT hardware designs, we propose a novel \textit{\textbf{MorphAtt}} accelerator with a streamlined architecture, comprising SpikeQKV, SpikeAtten, RepConv, and inter-module buffers.
Details of the architecture are discussed in the following sub-sections.

\subsection{Overall Architecture}

As illustrated in Figure~\ref{Fig_MorphAtt}(a), the main parts include three primary processing modules: SpikeQKV, SpikeAtten, and RepConv.
To alleviate the data traffic congestion of global on-chip memory access while maximizing spatial data reuse, specialized inter-module buffers (QKV\_Buffers and Data\_Buffer) are integrated into the execution flow, as shown in Figure~\ref{Fig_MorphAtt}(a).
Here, $H$, $N$, $D$, and $T$ denote the number of attention heads, input tokens, feature dimensions, and processing timesteps, respectively.

\subsection{Spiking Query-Key-Value Generator (SpikeQKV)}

This module is designed as one of the primary modules responsible for preparing the incoming feature maps and creating the standardized matrices for Query ($Q$), Key ($K$), and Value ($V$), which are necessary for attention operations; as shown in Figure~\ref{Fig_MorphAtt}(b).
SpikeQKV utilizes three different hardware pipelines for simultaneous treatment of separate matrix streams of Query ($Q$), Key ($K$), and Value ($V$).
In this SpikeQKV module, there are three basic building blocks, as described below.
\begin{itemize}
    \item \textbf{Spike Generator (SpikeGen):}
    This unit converts real-valued inputs into event-based representation in the form of binary spiking events (i.e., $0$ or $1$); see an overview in Figure~\ref{Fig_SpikeGen}. 
    The conversion is carried out by spiking neurons with the Leaky Integrate-and-Fire (LIF) model, expressed with the following.
    \begin{equation}
        U[t] = U[t-1] + \frac{1}{\tau} \big(x[t] - (U[t-1] - U_{rst}) \big)
    \end{equation}
    where $x[t]$ is a real-valued input token at timestep $t$, $U[t]$ is a neuron membrane potential at timestep $t$, $U_{rst}$ is a neuron reset potential, and $\tau$ is a membrane time constant.
    \begin{figure}[t]
        \centering
        \includegraphics[width=0.95\linewidth]{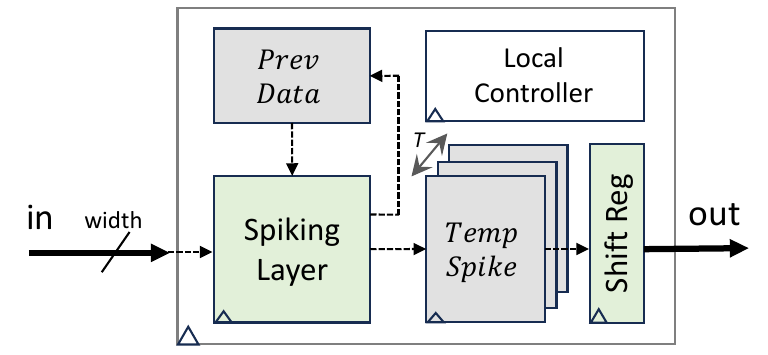}
        \caption{SpikeGen internal architecture.}
        \label{Fig_SpikeGen}
    \end{figure}
    To minimize hardware area and eliminate costly division units, we set the time constant $\tau = 2$. 
    Consequently, the scaling factor of $1/\tau$ can be easily translated to the logical right-shift operation ($\gg 1$), which significantly reduces combinational gate count and delays. Regarding the initial boundary conditions for the potential matrix in the first timestep $t = 0$, the history matrix $U(-1)$ is assumed to be of zero value only.
    Operations in SpikeGen include the following steps:
    \begin{itemize}
        \item \textit{Thresholding:} 
        When the potential value $U(t)$ reaches the specified threshold value ($U_{thr}$), a spike `1' is generated and placed in an internal temporary spike buffer (TempSpike); otherwise no-spike event `0' is recorded.
        \item \textit{History Tracking:} 
        The history buffer records the residual non-zero values of $U(t)$ to act as a baseline potential for subsequent timestep $t+1$.
        \item \textit{Alignment:} 
        The output data is shifted using an internal shift-register structure into the desired \textit{spiked\_img} buffering format. 
        Each row in the output is tightly packed into $H \times D$ bit streams with a fixed $T \times N \times (H \times D)$ dimension.
    \end{itemize}
    \item \textbf{RepConv within SpikeQKV:}
    Binary streams stored in the spiked\_img buffer are sent to the RepConv unit for processing sparse spikes along the channels of $Q$, $K$, $V$, and several heads, to compute local spatial dependencies.
    The outputs are then streamed to the $Q$/$K$/$V$-dedicated SpikeGen units to compute the unified head-concatenated spike matrices for $Q$, $K$, and $V$. 
    The outputs are then saved in the customized spiked\_img\_Q, spiked\_img\_K, and spiked\_img\_V buffers with fixed dimensions of $T \times N \times (H \times D)$ for their projection.
    \item \textbf{Head Separator:} 
    This unit aims to separate the head-concatenated spike matrices for each individual head using an efficient time-multiplexed routing mechanism, from which $Q$, $K$, and $V$ arrays can be extracted from the spike matrices. 
    This unit is controlled through two internal signals and employs two separate streams.
    \begin{itemize}
        \item It isolates the un-transposed $N \times D$ of $Q$ matrix through dynamic isolation of localized $D$-bit pieces per active token index position.
        \item It isolates and transposes on-the-fly the respective $D \times N$ configurations of $K$ and $V$ projection matrices to provide effective column-wise streaming.
    \end{itemize}
    The isolated matrices are then stored in QKV\_Buffers, encompassing Single-Head Query buffer (SH\_Q), Single-Head Key buffer (SH\_K), and Single-Head Value buffer (SH\_V).
    QKV\_Buffers hold the total capacity of $T \times (N \times D)$ for each $H$ independent heads for any projection type ($Q$, $K$, or $V$), as shown in Figure~\ref{Fig_MorphAtt}(c). 
\end{itemize}

\subsection{Spiking Multi-Head Self-Attention Engine (SpikeAtten)}

SpikeAtten module receives matrices of spiking activities from QKV\_Buffers for each head to execute MHSA operations.
As illustrated in Figure~\ref{Fig_MorphAtt}(d), the microarchitecture of the SpikeAtten module comprises three primary units, including \textit{Single-Head Self-Attention (SHSA)}, \textit{Concatenation (Concat)}, and \textit{SpikeGen}.
\begin{itemize}
    \item \textbf{Single-Head Self-Attention (SHSA):} 
    Self-attention operation typically involves calculating the attention score matrix based on sequence $(Q K^T) V$, which requires high computational complexity $O(N^2D)$.
    Since the number of tokens ($N$) in edge vision workloads is generally significantly larger than the feature size ($D$), such quadratic dependence creates heavy computational and memory access costs.
    To address this limitation, the SHSA rearranges the operation sequence as $Q (K^T V)$, leading to lower computational complexity with $O(ND^2)$. 
    For operational parameters of $N=196$ and $D=64$, such a right-to-left transformation results in a direct three-fold reduction in computational effort, regardless of any further optimization for sparsity.
    Moreover, given that all input data are in binary spikes, the unit is multiplier-free.
    SHSA unit employs bitwise operations through three different stages of computation as shown in Figure~\ref{Fig_SHSA}, which are described below.
    \begin{itemize}
        \item Step-1. \textit{$K^T V$ Calculation:} 
        Due to the binary nature of spikes in the $K$ and $V$ arrays, their dot product can be implemented as a bit-wise AND operation, followed by pop-count calculation. 
        The data in the internal Single-Head Key buffer (SH\_K) and Single-Head Value buffer (SH\_V) are simultaneously read row-wise, computed together with AND operation, whose results are aggregated and then saved as $K^T V$ matrices for $T$ timesteps. 
        \item Step-2. \textit{Gating Query Multiplication and Sleep Mode:} 
        Each row of SH\_Q is read into both static registers and into a shift register architecture. 
        Meanwhile, each row in the pre-computed $K^T V$ matrix is accumulated by the Accumulator dynamically based on the gating criterion. 
        Specifically, an addition is performed when the logical AND result of two signals meets the following conditions: (a) if the MSB of the Query shift register is `1', and (b) a reduction OR operation on the targeted $K^T V$ row indicates non-zero weights.
        If either signal resolves to zero (indicating the absence of active information), the Accumulator dynamically bypasses the execution cycle and enters the low-power ``\textit{sleep mode}'' to skip unnecessary switching power.
        Once the shift-register shifting finishes but the remaining $K^T V$ rows persist, the holding register reloads the original Query layout back into the shift register, iteratively repeating this sequence until both tracking domains are finished.
        \item Step 3. \textit{Attention Output Buffering}: The aggregated outputs that were obtained from each active head are sent to $T$ intermediate buffers, where each buffer is designed to accommodate an $N \times D$ size data block.
    \end{itemize}
    \begin{figure}[h]
    \centering
    \includegraphics[width=0.95\linewidth]{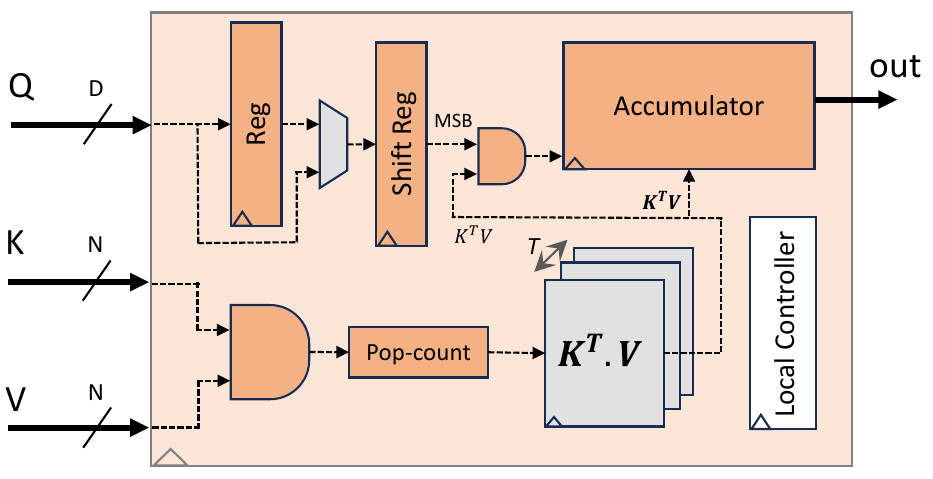}
    \caption{Single-Head Self-Attention (SHSA) architecture.}
    \label{Fig_SHSA}
    \end{figure}

    \item \textbf{Concatenation (Concat) Unit:} 
    After the self-attention operations are carried out on all the $H$ separate heads simultaneously, the Concat Unit integrates the processed information.
    Information produced by the active heads are sampled sequentially and then concatenated into one unique attention matrix buffer (BuffAtten) whose total size is $N \times (H \times D)$ over all considered timesteps $T$.
    It is achieved through our sophisticated time-multiplexed multiplexer network. 

    \item \textbf{Post-Attention SpikeGen Unit:} 
    Since the transformer pipeline follows the principles of an event-driven computation paradigm, it is crucial to ensure that the high-precision real values accumulated in BuffAtten can be converted back to spikes before being used further.
    That is why the information coming out of the buffer is further pipelined through the post-attention \textit{SpikeGen} unit that operates analogously to the low-complexity LIF engine by leveraging SpikingLayer as shown in Figure~\ref{Fig_SpikeGen}.
    The final spiking attention output will be written to the shared on-chip data buffer (Data\_Buffer) whose size is $T \times N \times (H \times D)$, as shown in Figure~\ref{Fig_MorphAtt}(e).
\end{itemize}

\subsection{\textbf{Reparameterization Convolution (RepConv)}}

To accommodate the reparameterization-based spatial features processing, we design the RepConv module as shown in Figure~\ref{Fig_MorphAtt}(f). 
The pipeline design in RepConv includes eight stages, including \textit{ReshapeIn}, \textit{SpikeConv1D}, and \textit{BatchNorm} (for Phase 1); \textit{Padding}, \textit{Conv2D}, \textit{Conv1D}, and \textit{BatchNorm} (for Phase 2); as well as \textit{ReshapeOut}. 
The functionalities of each stage are described below.
\begin{figure*}[t]
    \centering
    \includegraphics[width=\linewidth]{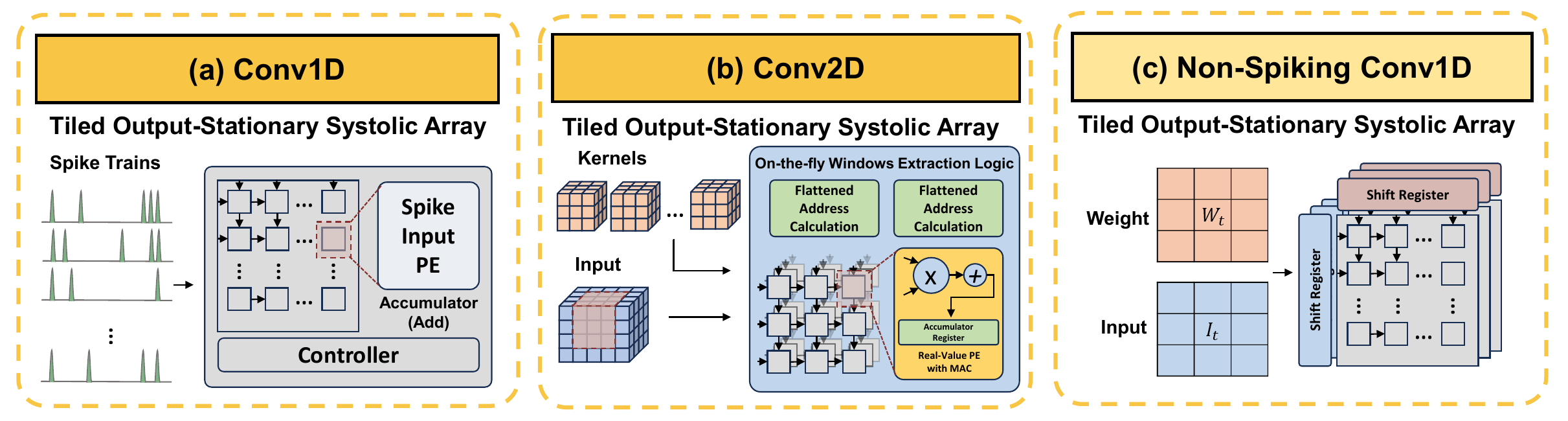}
    \vspace{-0.4cm}
    \caption{Overview of Convolutional Architectures in MorphAtt: \textbf{(a)} Conv1D, \textbf{(b)} Conv2D, and \textbf{(c)} Non-Spiking Conv1D.}
    \label{Fig_Convs}
    \vspace{-0.3cm}
\end{figure*}
\begin{itemize}
    \item \textbf{ReshapeIn}: 
    It reshapes multi-dimensional tensors into sequential token streams. 
    This module passes the serialized input data across temporal steps into an internal buffer.
    Tailored to a 3-stage finite-state machine (FSM), it captures a 2D slice of the input with $N \times (D \times H)$ dimension, and writes it to the internal buffer with the depth of $T$. 
    Addressing of $T$ depth is controlled by a dynamic timestep counter, counting from 0 to $T$-1. 
    Therefore, the result would be a 3D data format with $T \times N \times (D \times H)$ dimension.
    \item \textbf{Conv1D}: 
    It computes binary spike signals with conditional accumulations, and hence no hardware multipliers are needed; see Figure~\ref{Fig_Convs}(a). 
    This unit replaces the conventional convolution operation with a low-power spiking mechanism. 
    The processing core includes an FSM and a parameterized systolic array with $S \times S$ Processing elements (PE) designed for tiling across $N$ (spatial dimension) and $F$ (channel dimension). Each PE processes the input stream across $D \times H$. 
    In each clock cycle, the corresponding signed weight is added to the accumulator based on the input spike condition; otherwise, the accumulation is in an idle state to eliminate power consumption. 
    Tile division and execution are performed by a controller, and multi-time parallelism is supported across multiple systolic arrays as well.
    \item \textbf{Padding}: 
    It inserts padding data to fulfill spatial dimension constraints. 
    This unit aims to fit the data dimension to the subsequent units. 
    It performs zero or constant value padding across channels and over time. 
    Based on a sequence of $T$ timesteps, the flattened input feature size of $\sqrt{N} \times \sqrt{N}$ is mapped to a 2D grid with side length $\sqrt{N} + 2\times Pad$, where $Pad$ is the padding value.
    \item \textbf{Conv2D}: 
    It conducts $3\times3$ spatial convolution through the output-stationary systolic architecture with on-the-fly sliding window extraction; see Figure~\ref{Fig_Convs}(b). 
    This unit performs a $3 \times 3$ 2D convolution operation across temporal dimension $T$ and channel dimension $F$. 
    An FSM is responsible to choose the respective index for $3 \times 3$ window selection based on the combinatorial indexing as follows: $W_{m,n} = input\_stream[(r+m) \sqrt{N}+(c+n)] \quad$ for $m,n \in \{0,1,2\}$, where $r$ and $c$ denote the current row and column index, respectively. 
    The extracted window and the $3 \times 3$ kernel are fed to PEs, each having nine Multiply-Accumulate (MAC) units.
    \item \textbf{Non-Spiking Conv1D}: 
    It implements channel-wise projection using the tiled systolic architecture and time-multiplexed dataflow; transposition operation is done on-the-fly by the routing fabric; see Figure~\ref{Fig_Convs}(c). 
    This unit performs time-multiplexed channel-wise projection using tiled systolic arrays across timesteps. 
    To avoid pre-processing latency caused by the feature tensor, transposition ($D \times N \rightarrow N \times D$) is performed on-the-fly. 
    Moreover, two groups of Shift Registers achieve data alignment appropriately and feed them to the $S \times S$ grid of PEs.
    \item \textbf{BatchNorm (Phases 1 \& 2)}: 
    It normalizes the intermediate tensor through scaling and offset. 
    This unit performs feature normalization across the input dimension of $T \times N \times F$. 
    To eliminate hardware load and resource complexity from division and square-root operation~\cite{Ref_Putra_RegFreeSqrt_IC3INA14} during the inference phase, inverse standard deviation parameter $ inv\_std = 1/\sqrt{\sigma^2 + \epsilon}$ is computed offline. 
    The inputs are then treated as Q4.12 fixed-point numbers, and the normalization, scaling, and shifting process are performed according to: $output[p][i][j] = ((input[p][i][j] - \mu_{p,j}) \cdot inv\_std_{p,j} \cdot \gamma_{p,j}) + \beta_{p,j}$.
    \item \textbf{ReshapeOut}: 
    It converts processed tokens back to a multi-dimensional matrix format. 
    To achieve this, the sequence data are reformatted to multi-dimensional tensors using the ReshapeOut unit, whose size dimension is $T \times N \times F$.
\end{itemize}

\subsection{Processing Dataflow}

To enable efficient MHSA processing, we employ streamlined processing that subsequently propagates data across cascaded modules.
Specifically, we devise two dataflow variants by leveraging the inter-module buffers for holding intermediate results, including: (1) \textit{sequential-based processing} that computes an input sample at a time to minimize power cost, and (2) \textit{pipeline-based processing} that computes multiple input samples at a time to improve the throughput.

\begin{table*}[t]
\caption{Comparison against the state-of-the-art digital SViT accelerators.}
\centering
\scriptsize
\begin{tabular}{|c|c|c|c|c|c|c|c|c|} 
\hline 
\textbf{Evaluation} & 
\begin{tabular}[c]{@{}c@{}} \textbf{Design of} \\ \cite{Ref_Li_SViTacc_APACE24} \end{tabular} & 
\begin{tabular}[c]{@{}c@{}} \textbf{SpikeTA} \\ \cite{Ref_Gao_SpikeTA_TCAD25} \end{tabular} & 
\begin{tabular}[c]{@{}c@{}} \textbf{FireFly-T} \\ \cite{Ref_Li_FireFlyT_arXiv25} \end{tabular} &
\begin{tabular}[c]{@{}c@{}} \textbf{VESTA} \\ \cite{Ref_Chen_VESTA_APCCAS24} \end{tabular} & 
\begin{tabular}[c]{@{}c@{}} \textbf{Bishop} \\ \cite{Ref_Xu_Bishop_ISCA25} \end{tabular} & 
\begin{tabular}[c]{@{}c@{}} \textbf{Design of} \\ \cite{Ref_Fang_EMAfree_TCASAI25} \end{tabular} & 
\begin{tabular}[c]{@{}c@{}} \textbf{Design of} \\  \cite{Ref_Xu_SThw3D_ICCAD24} \end{tabular} & \begin{tabular}[c]{@{}c@{}} \textbf{\textit{MorphAtt}} \\ (ours) \end{tabular} \\
\hline 
\hline
Paradigm & \begin{tabular}[c]{@{}c@{}} 2D \end{tabular} & \begin{tabular}[c]{@{}c@{}} 2D \end{tabular} & \begin{tabular}[c]{@{}c@{}} 2D \end{tabular} & \begin{tabular}[c]{@{}c@{}} 2D \end{tabular} & \begin{tabular}[c]{@{}c@{}} 2D \end{tabular} & \begin{tabular}[c]{@{}c@{}} 2D \end{tabular} & \begin{tabular}[c]{@{}c@{}} 3D \end{tabular} & \begin{tabular}[c]{@{}c@{}} 2D \end{tabular} \\ 
\hline
\begin{tabular}[c]{@{}c@{}} Technology \\ (nm) \end{tabular} & FPGA  & FPGA & FPGA & 28 & 28 & 28 & 28 & 32 \\ 
\hline
\begin{tabular}[c]{@{}c@{}} Precision \\ (bit) \end{tabular} & 10 & 8 & 4 & 8 & multi-bit & 8 & 8 & 8 \\ 
\hline
Area (mm$^2$) & \textcolor{gray}{N/A} & \textcolor{gray}{N/A} & \textcolor{gray}{N/A} & 0.84 & 2.96 & 6.5 & \begin{tabular}[c]{@{}c@{}} 0.20 \end{tabular} & 1.49 \\
\hline
Power (mW) & $\sim$12000 & $\sim$71759 & $\sim$4349 & 416 & 627 & 222.65 & \begin{tabular}[c]{@{}c@{}} 201.63 \end{tabular} & 38.96 - 55.12 \\ 
\hline
\begin{tabular}[c]{@{}c@{}} Frequency \\ (MHz) \end{tabular} & 200 & 450 & 300 & 500 & 500 & 200 & 500 & 100 \\
\hline
\begin{tabular}[c]{@{}c@{}} Throughput \\ (GOPS) \end{tabular} & 307.2 & 28990 & 3397 & 4096 & \textcolor{gray}{N/A} & $\sim$6100 & \textcolor{gray}{N/A} & 792.6 - 1605\\
\hline
\begin{tabular}[c]{@{}c@{}} Efficiency \\ (GOPS/W) \end{tabular} & 25.6 & 403.99 & 781.13 & 9.84K & \textcolor{gray}{N/A} & 23.5K - 27.9K & \textcolor{gray}{N/A} & 20.33K - 29.12K\\ 
\hline
\end{tabular}
\label{Table_CompareSoA}
\end{table*}

\section{Evaluation Methodology}
\label{Sec_Eval}

To evaluate our MorphAtt accelerator design, we first implement the design in Register-Transfer Level (RTL), and then perform a comprehensive experimental evaluation using the ASIC design flow.
For performance analysis, we synthesize the MorphAtt accelerator design using the Synopsys Design Compiler with a standard cell library with 32nm CMOS technology. 
The timing constraint value is set to 100 MHz.
Post-synthesis gate-level simulations are performed to obtain switching activity, which is then used to obtain the power consumption of MorphAtt. 
The toggle rates are extracted from simulations of the state-of-the-art Spike-Driven Transformer v2 (SDTv2) model~\cite{Ref_Yao_SpikeDrivenTransformer2_ICLR24} running the CIFAR-100 dataset. 
The SDTv2 configuration includes $H=8$ attention heads, $N=196$ tokens, $D=64$ embedding dimensions, and $T=4$ timesteps following a configuration known to achieve competitive accuracy for vision tasks~\cite{Ref_Putra_QSViT_IJCNN25}.
We evaluate the MorphAtt design using several metrics. 
First is area (mm$^2$), obtained from the synthesis process.
Second is throughput (GOPS), obtained by calculating the computation graph considering the SDTv2 model with CIFAR-100 workload.
Third is energy efficiency (TOPS/W), obtained through a calculation that leverages the throughput and total power consumption of the design.

\section{Results and Discussion}
\label{Sec_Results}
 
\subsection{Performance and Energy-Efficiency Analysis}

Experimental results for performance and energy-efficiency are summarized and compared to the state-of-the-art digital SViT accelerators in Table~\ref{Table_CompareSoA}.
These results show that FPGA-based solutions~\cite{Ref_Gao_SpikeTA_TCAD25, Ref_Li_FireFlyT_arXiv25, Ref_Li_SViTacc_APACE24} offer diverse performance benefits depending on their design architectures and dataflow patterns, ranging between 300 GOPS to 29 TOPS. 
However, they incur high power consumption (i.e., more than 4W), thereby leading to less than 1 TOPS/W of energy efficiency. 
The reason is that, flexibility in FPGA platforms may introduce notable overheads in routing capacitance, parasitic resistance, and switching activities, therby limiting their energy efficiency gains. 

These results also show that ASIC-based solutions~\cite{Ref_Chen_VESTA_APCCAS24, Ref_Fang_EMAfree_TCASAI25, Ref_Xu_SThw3D_ICCAD24, Ref_Xu_Bishop_ISCA25} often achieve higher performance benefits than FPGA-based solutions by offering at least 4 TOPS of throughput, while incurring less than 1 Watt of power consumption with diverse area sizes, ranging from 0.2 mm$^2$ to 6.5 mm$^2$.
These conditions enable up to 27.9 TOPS/W of energy efficiency, demonstrating their capabilities in achieving higher performance and higher efficiency gains than FPGA-based solutions.
The reason is that, ASIC-based designs are tailored to specific functionalities with minimal components and optimized routing, thereby substantially minimizing the latency, area, and power/energy consumption.

Meanwhile, our MorphAtt accelerator achieves 792 GOPS of throughput, while incurring about 38.96 mW of power consumption for sequential-based processing, thereby leading to about 20.3 TOPS/W of energy efficiency. 
It also improves the performance further to 1605 GOPS of throughput, while incurring about 55.12 mW of power consumption for pipeline-based processing, thereby leading to about 29.1 TOPS/W of energy efficiency. 
MorphAtt occupies 1.5 mm$^2$ of area, which is relatively small for ASIC-based solutions. 
These results indicate that, our MorphAtt offers better performance and energy-efficiency trade-offs as compared to the state-of-the-art from both FPGA- and ASIC-based solutions, by offering competitive performance with ultra-low power consumption.
These strong performance and energy-efficiency offered by our MorphAtt accelerator are attributed to the following reasons. 
\begin{itemize}
    \item Each MorphAtt module is designed from latency-, area-, and power-optimized architectures for components/units, modules, and buffers to minimize overall costs.
    \item The cascaded modules and pipelined dataflow in MorphAtt enable high throughput.
\end{itemize}
Overall, these results also highlight that our MorphAtt accelerator is suitable for accelerating SViTs on power-constrained vision-based systems, which is especially important for diverse edge-AI computing use-cases. 

\subsection{MorphAtt Module Synthesis and Power Analysis}

\begin{table}[t]
\centering
\caption{Post-synthesis breakdown of MorphAtt modules (32nm HVT, 1.05V, 100 MHz).}
\label{Table_ModuleBreakdown}
\footnotesize
\begin{tabular}{|l|c|c|c|}
\hline
& \textbf{Total} & \textbf{Total} & \textbf{Power} \\
\textbf{Module} & \textbf{Area} & \textbf{Power} & \textbf{Distribution} \\
& (10$^3$ $\times$ $\mu$m$^2$) & (mW) & (\% of total) \\
\hline
\hline
SpikeQKV & 686.0 & 20.095 & 51.6\% \\
SpikeAtten & 9.4 & 0.209 & 0.6\% \\
RepConv & 42.6 & 1.267 & 3.2\% \\
On-chip Memories & 758.3 & 17.394 & 44.6\% \\
\hline
\textbf{Total} & \textbf{1496.3} & \textbf{38.965} & 100\% \\
\hline
\end{tabular}
\end{table}

\begin{figure}[t]
    \centering
    \includegraphics[width=\linewidth]{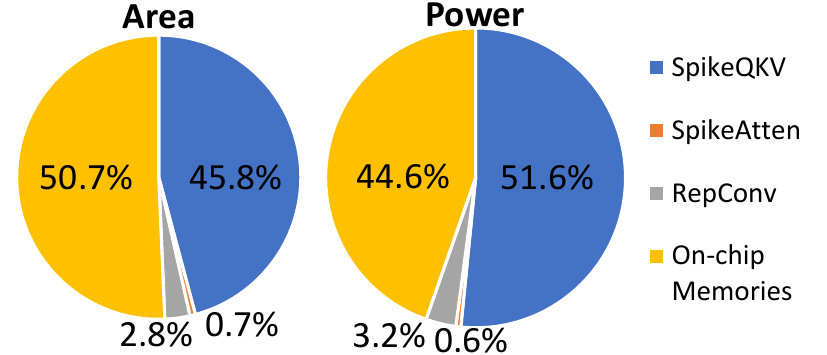}
    \caption{Pie-chart breakdown of MorphAtt for \textbf{(a)} area and \textbf{(b)} power consumption, across different modules.}
    \label{Fig_PieCharts}
    \end{figure}
    
We further analyze the performance benefits of MorphAtt design with respect to its post-synthesis breakdown across different MorphAtt modules; see Table~\ref{Table_ModuleBreakdown} and Figure~\ref{Fig_PieCharts}.
From these results, we make the following key observations.
\begin{itemize}
    \item SpikeQKV module has a data input latency of 5.35 ns at 100 MHz clock speed (10 ns clock period), giving us 4.61 ns of slack, indicating that the LIF design operates within 15\% of the theoretical timing limit for this technology node (i.e., 85\% utilization of the timing budget).
    \item 
    SpikeAtten module is highly efficient, consuming only 209 $\mu$W, which translates to a 96\% power reduction compared to regular attention modules using multipliers (i.e., about 5.2 mW with 8-bit precision).
    This high efficiency gain can be attributed to two main factors. First, redesigning the computation in attention operations optimizes the computation complexity from $\mathcal{O}(N^2D)$ to $\mathcal{O}(ND^2)$. 
    Second, eliminating all multiplications and using AND+popcount helps optimize per-operation energy from $\sim$500 fJ to $\sim$12 fJ in 32nm CMOS technology.
    \item 
    In contrast to previous SViT hardware accelerators without RepConv capabilities, MorphAtt incorporates these functionalities at an extra cost of just 1.267 mW (which is only 3.3\% of total power consumption) and 42.6 k $\mu$m$^2$ of area overhead, thereby providing low-cost support for processing reparameterization in SViTs.
    \item 
    The architecture provides a balanced ratio of 49\% compute to 51\% memory area. 
    This is due to the dataflow-driven streamlined design where compute modules handle as much data as the inter-module buffers can hold, thereby minimizing accesses to off-chip DRAMs and enabling high performance of 792 GOPS.
    \item 
    Each MorphAtt module meets timing requirements with slack, where RepConv and SHSA modules have a slack of 7.26 ns - 7.49 ns, which implies that these modules can run up to 135 MHz without any modification to their design.
    This margin provides a potential to increase throughput by 35\% (i.e., up to 1,070 GOPS). 
    \item Although the memory macros account for 51\% of the entire chip area, the SpikeQKV module (i.e., LIF-based spike generation and head separation circuitry) accounts for 51.6\% of total power consumption (20.1 mW). 
    \item 
    The above-discussed results are obtained considering the worst-case scenario in terms of temperature (125°C) and HVT cells. 
    The total power consumption in MorphAtt is expected to be reduced by 35\%-50\% at operating temperatures (25°C-85°C), due to the exponential leakage-versus-temperature curve, resulting in 20-25 mW effective power consumption.
\end{itemize}

\subsection{Further Discussion on Processing Time Diagram}

\begin{figure}[t]
    \centering
    \includegraphics[width=\linewidth]{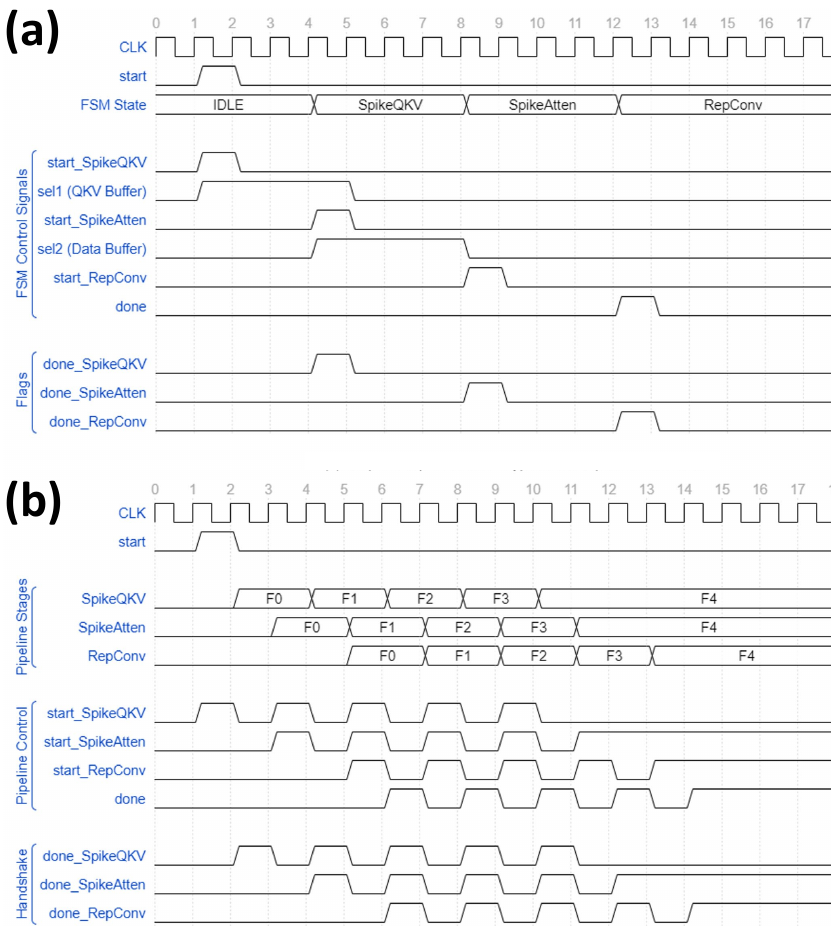}
    \caption{\textbf{(a)} The timing diagram of the main FSM for sequential-based processing in MorphAtt. \textbf{(b)} The hypothetical timing diagram of the main FSM for pipeline-based processing in MorphAtt.}
    \label{Fig_Top_Timing}
\end{figure}

To demonstrate the top-level execution timeline of our MorphAtt, the timing diagram of the main FSM for sequential-based processing is shown in Figure~\ref{Fig_Top_Timing}(a), while the hypothetical timing diagram of the main FSM for pipeline-based processing is shown in Figure~\ref{Fig_Top_Timing}(b).
By asserting the start signal ($start$), the SpikeQKV module starts generating $Q$, $K$, and $V$ matrices, then stores them in the intermediate buffers (QKV\_Buffers). 
Once the process is done, the completion signal ($done\_SpikeQKV$) is generated. 
Based on the selection signals, intermediate buffers start routing the data to the SpikeAtten module, which is activated by its activation signal ($start\_SpikeAtten$). 
After the attention operations, the results are written to Data\_Buffer and its done signal ($done\_SpikeAtten$) is asserted. 
Eventually, the FSM toggles the selection for Data\_Buffer in order to feed the data to the RepConv module. 
Once reparameterization operations are finished, the global done signal ($done$) is asserted as well, returning the system to IDLE state and ready for a new data sequence.

\section{Conclusion}
\label{Sec_Conclude}

In this work, we propose \textit{\textbf{MorphAtt}}, a novel low-power digital accelerator for SViT inference acceleration through streamlined processing.
The key idea behind MorphAtt design is the cascaded architecture with SpikeQKV, SpikeAtten, and RepConv modules, as well as inter-module buffers (i.e., QKV\_Buffers and Data\_Buffer) to mitigate traffic congestion in on-chip memory accesses.  
Under synthesis using 32nm CMOS technology, the MorphAtt accelerator operates in the range of 792-1605 GOPS of throughput with 38.96-55.12 mW of power and 1.49 mm² chip area, thereby offering energy efficiency up to 29.12 TOPS/W.
These results also demonstrate that our MorphAtt offers better performance and energy-efficiency trade-offs compared to state-of-the-art, thereby making it suitable for enabling highly energy-efficient vision-based AI systems at the edge, such as autonomous agents and mobile robots.



\bibliography{aaai2027} 

@INPROCEEDINGS{Ref_Putra_RegFreeSqrt_IC3INA14,
  author={Putra, Rachmad Vidya Wicaksana and Adiono, Trio},
  booktitle={2014 International Conference on Computer, Control, Informatics and Its Applications (IC3INA)}, 
  title={A register-free and homogenous architecture for square root algorithm}, 
  year={2014},
  volume={},
  number={},
  pages={64-68},
  doi={10.1109/IC3INA.2014.7042602}}

@inproceedings{Ref_Shi_SpikingResformer_CVPR24,
  title={SpikingResformer: Bridging ResNet and Vision Transformer in Spiking Neural Networks},
  author={Shi, Xinyu and Hao, Zecheng and Yu, Zhaofei},
  booktitle={Proceedings of the IEEE/CVF Conference on Computer Vision and Pattern Recognition (CVPR)},
  pages={5610--5619},
  year={2024}
}

@article{Ref_Fang_EMAfree_TCASAI25,
  title={A 28nm Spiking Vision Transformer Accelerator with Dual-Path Sparse Compute Core and EMA-free Self-Attention Engine for Embodied Intelligence},
  author={Fang, Chaoming and Shen, Ziyang and Li, Tianyang and Zhao, Shiqi and Tian, Fengshi and Yang, Jie and Sawan, Mohamad},
  journal={IEEE Transactions on Circuits and Systems for Artificial Intelligence (TCASAI)},
  year={2025},
  publisher={IEEE}
}

@inproceedings{Ref_Xu_SThw3D_ICCAD24,
  author={Xu, Boxun and others},
  booktitle={2024 ACM/IEEE International Conference On Computer Aided Design (ICCAD)}, 
  title={Spiking Transformer Hardware Accelerators in 3D Integration}, 
  year={2024},
  volume={},
  number={},
  pages={1-9},
  doi={10.1145/3676536.3676826}}

@inproceedings{Ref_Xu_Bishop_ISCA25,
  title={Bishop: Sparsified bundling spiking transformers on heterogeneous cores with error-constrained pruning},
  author={Xu, Boxun and others},
  booktitle={Proceedings of the 52nd Annual International Symposium on Computer Architecture},
  pages={944--957},
  year={2025}
}

@article{Ref_Li_FireFlyT_arXiv25,
  title={FireFly-T: High-Throughput Sparsity Exploitation for Spiking Transformer Acceleration with Dual-Engine Overlay Architecture},
  author={Li, Tenglong and others},
  journal={arXiv preprint arXiv:2505.12771},
  year={2025}
}

@article{Ref_Gao_SpikeTA_TCAD25,
  title={Advancing neuromorphic architecture towards emerging spiking neural network on fpga},
  author={Gao, Yingxue and others},
  journal={IEEE Transactions on Computer-Aided Design of Integrated Circuits and Systems (TCAD)},
  year={2025},
  publisher={IEEE}
}

@inproceedings{Ref_Chen_VESTA_APCCAS24,
  title={VESTA: A Versatile SNN-Based Transformer Accelerator with Unified PEs for Multiple Computational Layers},
  author={Chen, Ching-Yao and others},
  booktitle={2024 IEEE Asia Pacific Conference on Circuits and Systems (APCCAS)},
  organization={IEEE},
  pages={6--10},
  year={2024},
}

@inproceedings{Ref_Li_SViTacc_APACE24,
  title={An efficient sparse hardware accelerator for spike-driven transformer},
  author={Li, Zhengke and Mao, Wendong and Zhang, Siyu and Dong, Qiwei and Wang, Zhongfeng},
  booktitle={2024 IEEE Asia-Pacific Conference on Applied Electromagnetics (APACE)},
  pages={250--253},
  year={2024},
  organization={IEEE}
}

@inproceedings{Ref_Touvron_TrainingDeIT_ICML21,
  title={Training data-efficient image transformers \& distillation through attention},
  author={Touvron, Hugo and Cord, Matthieu and Douze, Matthijs and Massa, Francisco and Sablayrolles, Alexandre and J{\'e}gou, Herv{\'e}},
  booktitle={International Conference on Machine Learning (ICML)},
  pages={10347--10357},
  year={2021},
}

@INPROCEEDINGS{Ref_Putra_SNNonCNP_IJCNN25,
  author={Putra, Rachmad Vidya Wicaksana and Wickramasinghe, Pasindu and Shafique, Muhammad},
  booktitle={2025 International Joint Conference on Neural Networks (IJCNN)}, 
  title={Enabling Efficient Processing of Spiking Neural Networks with On-Chip Learning on Commodity Neuromorphic Processors for Edge AI Systems}, 
  year={2025},
  volume={},
  number={},
  pages={1-8},
  doi={10.1109/IJCNN64981.2025.11228298}}

@INPROCEEDINGS{Ref_Putra_QSViT_IJCNN25,
  author={Putra, Rachmad Vidya Wicaksana and Iftikhar, Saad and Shafique, Muhammad},
  booktitle={2025 International Joint Conference on Neural Networks (IJCNN)}, 
  doi={10.1109/IJCNN64981.2025.11228251},
  title={QSViT: A Methodology for Quantizing Spiking Vision Transformers}, 
  year={2025},
  volume={},
  number={},
  pages={1-8},
  }

@article{roy2019towards,
  title={Towards spike-based machine intelligence with neuromorphic computing},
  author={Roy, Kaushik and Jaiswal, Akhilesh and Panda, Priyadarshini},
  journal={Nature},
  volume={575},
  number={7784},
  pages={607--617},
  year={2019},
  publisher={Nature Publishing Group UK London}
}

@ARTICLE{Ref_Putra_SpikeNAS_TAI25,
  author={Putra, Rachmad Vidya Wicaksana and Shafique, Muhammad},
  journal={IEEE Transactions on Artificial Intelligence (TAI)}, 
  title={SpikeNAS: A Fast Memory-Aware Neural Architecture Search Framework for Spiking Neural Network-based Embedded AI Systems}, 
  year={2025},
  volume={},
  number={},
  pages={1-12},
  doi={10.1109/TAI.2025.3586238}}

@inproceedings{Ref_Zhou_Spikformer_ICLR23,
title={Spikformer: When Spiking Neural Network Meets Transformer },
author={Zhaokun Zhou and others},
booktitle={11th International Conference on Learning Representations (ICLR)},
year={2023},
}

@inproceedings{
Ref_Yao_SpikeDrivenTransformer2_ICLR24,
title={Spike-driven Transformer V2: Meta Spiking Neural Network Architecture Inspiring the Design of Next-generation Neuromorphic Chips},
author={Man Yao and others},
booktitle={The 12th International Conference on Learning Representations (ICLR)},
year={2024},
}

@inproceedings{Ref_Yao_SpikeDrivenTransformer_NeurIPS23,
title={Spike-driven Transformer},
author={Man Yao and others},
booktitle={The 37th Conference on Neural Information Processing Systems (NeurIPS)},
year={2023},
}

@article{Ref_Khan_SurveyViT_CSUR22,
  title={Transformers in Vision: A Survey},
  author={Khan, Salman and Naseer, Muzammal and Hayat, Munawar and Zamir, Syed Waqas and Khan, Fahad Shahbaz and Shah, Mubarak},
  journal={ACM Computing Surveys (CSUR)},
  year={2022},
  number={10s},
  pages={1--41},
  volume={54},
  publisher={ACM New York, NY}
}

@article{Ref_Han_SurveyViT_TPAMI22,
  title={A Survey on Vision Transformer},
  author={Han, Kai and others},
  journal={IEEE Transactions on Pattern Analysis and Machine Intelligence (TPAMI)},
  volume={45},
  number={1},
  pages={87--110},
  year={2022},
  publisher={IEEE}
}

@inproceedings{Ref_Dosovitskiy_Transformers_ICLR21,
title={An Image is Worth 16x16 Words: Transformers for Image Recognition at Scale},
author={Alexey Dosovitskiy and others},
booktitle={International Conference on Learning Representations (ICLR)},
year={2021},
url={https://openreview.net/forum?id=YicbFdNTTy}
}

@article{Ref_Rathi_SNNsurvey_CSUR23,
author = {Rathi, Nitin and Chakraborty, Indranil and Kosta, Adarsh and Sengupta, Abhronil and Ankit, Aayush and Panda, Priyadarshini and Roy, Kaushik},
title = {Exploring Neuromorphic Computing Based on Spiking Neural Networks: Algorithms to Hardware},
year = {2023},
issue_date = {December 2023},
publisher = {Association for Computing Machinery},
address = {New York, NY, USA},
volume = {55},
number = {12},
url = {https://doi.org/10.1145/3571155},
doi = {10.1145/3571155},
issn = {0360-0300},
journal = {ACM Computing Surveys},
articleno = {243}
}

@ARTICLE{Ref_Putra_FSpiNN_TCAD20,
  author={R. V. W. {Putra} and M. {Shafique}},
  journal={IEEE Transactions on Computer-Aided Design of Integrated Circuits and Systems (TCAD)},
  title={FSpiNN: An Optimization Framework for Memory-Efficient and Energy-Efficient Spiking Neural Networks}, 
  year={2020},
  volume={39},
  number={11},
  pages={3601-3613}}

@ARTICLE{Ref_Akopyan_TrueNorth_TCAD15,
author="F. {Akopyan} and J. {Sawada} and A. {Cassidy} and R. {Alvarez-Icaza} and J. {Arthur} and P. {Merolla} and N. {Imam} and Y. {Nakamura} and P. {Datta} and G. {Nam} and B. {Taba} and M. {Beakes} and B. {Brezzo} and J. B. {Kuang} and R. {Manohar} and W. P. {Risk} and B. {Jackson} and D. S. {Modha}",
journal="IEEE Transactions on Computer-Aided Design of Integrated Circuits and Systems (TCAD)",
volume="34",
number="10",
pages="1537-1557",
title="TrueNorth: Design and Tool Flow of a 65 mW 1 Million Neuron Programmable Neurosynaptic Chip",
month="Oct",
year="2015",
ISSN="",}

@ARTICLE{Ref_Davies_Loihi_MM18,
author="M. {Davies} and N. {Srinivasa} and T. {Lin} and G. {Chinya} and Y. {Cao} and S. H. {Choday} and G. {Dimou} and P. {Joshi} and N. {Imam} and S. {Jain} and Y. {Liao} and C. {Lin} and A. {Lines} and R. {Liu} and D. {Mathaikutty} and S. {McCoy} and A. {Paul} and J. {Tse} and G. {Venkataramanan} and Y. {Weng} and A. {Wild} and Y. {Yang} and H. {Wang}",
journal="IEEE Micro",
title="Loihi: A Neuromorphic Manycore Processor with On-Chip Learning",
year="2018",
volume="38",
number="1",
pages="82--99",
ISSN="",
month="Jan",}

@article{Ref_Maass_SNN_NeuNet97,
title = "Networks of spiking neurons: The third generation of neural network models",
journal = "Neural Networks",
volume = "10",
number = "9",
pages = "1659--1671",
year = "1997",
issn = "0893-6080",
author = "W. Maass",}

@ARTICLE{Ref_Pfeiffer_DLSNN_FNINS18,
AUTHOR="Pfeiffer, M. and Pfeil, T.",   
TITLE="Deep Learning With Spiking Neurons: Opportunities and Challenges",    
JOURNAL="Frontiers in Neuroscience",      
VOLUME="12",      
YEAR="2018",     
URL="",     
DOI="10.3389/fnins.2018.00774",     
ISSN="1662-453X",}

@article{Ref_Tavanaei_DLSNN_Neunet18,
title = "Deep learning in spiking neural networks",
journal = "Neural Networks",
volume = "111",
pages = "47--63",
year = "2019",
issn = "0893-6080",
doi = "https://doi.org/10.1016/j.neunet.2018.12.002",
author = "Amirhossein Tavanaei and Masoud Ghodrati and Saeed Reza Kheradpisheh and Timothée Masquelier and Anthony Maida",}

@INPROCEEDINGS{Ref_Putra_ReSpawn_ICCAD21,
  author={Putra, Rachmad Vidya Wicaksana and Hanif, Muhammad Abdullah and Shafique, Muhammad},
  booktitle={2021 IEEE/ACM International Conference On Computer Aided Design (ICCAD)}, 
  title={ReSpawn: Energy-Efficient Fault-Tolerance for Spiking Neural Networks considering Unreliable Memories}, 
  year={2021},
  volume={},
  number={},
  pages={1-9},
  doi={10.1109/ICCAD51958.2021.9643524}}

@article{Ref_Putra_RescueSNN_FNINS23,
  title={RescueSNN: enabling reliable executions on spiking neural network accelerators under permanent faults},
  author={Putra, Rachmad Vidya Wicaksana and Hanif, Muhammad Abdullah and Shafique, Muhammad},
  journal={Frontiers in Neuroscience (FNINS)},
  volume={17},
  pages={1159440},
  year={2023},
  publisher={Frontiers Media SA}
}

@inproceedings{Ref_Putra_SoftSNN_DAC22,
  title={SoftSNN: Low-Cost Fault Tolerance for Spiking Neural Network Accelerators under Soft Errors},
  author={Putra, Rachmad Vidya Wicaksana and Hanif, Muhammad Abdullah and Shafique, Muhammad},
  booktitle={59th ACM/IEEE Design Automation Conference (DAC)},
  pages={151--156},
  year={2022}
}


\end{document}